\documentclass{elsart}
 \usepackage{epsfig}

\usepackage{amssymb}

\begin{document}

\begin{frontmatter}


\title{Measurement of the relative  branching ratio BR(\boldmath$\Xi_c^+\to p^+ K^- \pi^+$)/BR(\boldmath$\Xi_c^+\to  \Xi^- \pi^+ \pi ^+$)
}

The FOCUS Collaboration

\author[ucd]{J.~M.~Link}
\author[ucd]{M.~Reyes\thanksref{atmichoacana}}
\author[ucd]{P.~M.~Yager}
\author[cbpf]{J.~C.~Anjos}
\author[cbpf]{I.~Bediaga}
\author[cbpf]{C.~G\"obel\thanksref{aturuguay}}
\author[cbpf]{J.~Magnin}
\author[cbpf]{A.~Massafferri}
\author[cbpf]{J.~M.~de~Miranda}
\author[cbpf]{I.~M.~Pepe\thanksref{atbahia}}
% ...

\address[ucd]{University of California, Davis, CA 95616} 
\address[cbpf]{Centro Brasileiro de Pesquisas F\'isicas, Rio de Janeiro, RJ, Brasil} 
\address[cinv]{CINVESTAV, 07000 M\'exico City, DF, Mexico} 
\address[cu]{University of Colorado, Boulder, CO 80309} 
\address[fnal]{Fermi National Accelerator Laboratory, Batavia, IL 60510} 
\address[fras]{Laboratori Nazionali di Frascati dell'INFN, Frascati, Italy I-00044} 
\address[ui]{University of Illinois, Urbana-Champaign, IL 61801} 
\address[iu]{Indiana University, Bloomington, IN 47405} 
\address[korea]{Korea University, Seoul, Korea 136-701} 
\address[milan]{INFN and University of Milano, Milano, Italy} 
\address[nc]{University of North Carolina, Asheville, NC 28804} 
\address[pavia]{Dipartimento di Fisica Nucleare e Teorica and INFN, Pavia, Italy} 
\address[pr]{University of Puerto Rico, Mayaguez, PR 00681} 
\address[sc]{University of South Carolina, Columbia, SC 29208} 
\address[ut]{University of Tennessee, Knoxville, TN 37996} 
\address[vu]{Vanderbilt University, Nashville, TN 37235} 
\address[wisc]{University of Wisconsin, Madison, WI 53706} 

\thanks[atmichoacana]{Present Address: Instituto de Fisica y Matematicas, Universidad Michoacana de San Nicolas de Hidalgo, Morelia, Mich., Mexico 58040} 
\thanks[aturuguay]{Present Address: Instituto de F\'isica, Faculdad de Ingenier\'i a, Univ. de la Rep\'ublica, Montevideo, Uruguay} 
\thanks[atbahia]{Present Address: Instituto de F\'isica, Universidade Federal da Bahia, Salvador, Brazil} 
\thanks[atmilan]{Present Address: INFN sezione di Milano, Milano, Italy} 
\thanks[atpueblo]{Present Address: Instituto de F\'{\i}sica, Universidad Aut\'onoma de Puebla,} 
\thanks[atncar]{Present Address: National Center for Atmospheric Research, Boulder, CO} 
\thanks[atvu]{Present Address: Vanderbilt University, Nashville, TN 37235} 
\thanks[atkriss]{Present Address:  Korea Research Institute of Standards and Science, P.O.Box 102, Yusong-Ku, Taejon 305-600, South Korea} 
\thanks[atcu]{Present Address: University of Colorado, Boulder 80309} 
\thanks[atbrescia]{Present Address: Dipartimento di Chimica e Fisica per l'Ingegneria e per i Materiali, Universit\'a di Brescia and INFN sezione di Pavia} 
\thanks[atbucharest]{Present Address: Nat. Inst. of Phys and Nucl. Eng., Bucharest, Romania} 
\thanks[atlucent]{Present Address: Lucent Technology} 
\thanks[atadamst]{Present Address: Adams State College, Alamosa, CO 81102}

\begin{abstract}
We report the observation of the Cabibbo suppressed decay
$\Xi_c^+ \to p K^-\pi^+$ using data collected with the FOCUS spectrometer
during the 1996--97
Fermilab fixed target run.
We find a $\Xi_c^+$ signal peak of $202\pm35$ events.
We have measured the relative branching ratios
BR$(\Xi^+_c\to p K^-\pi^+)/$BR$(\Xi^+_c\to\Xi^-\pi^+\pi^+)
= 0.234 \pm 0.047 \pm 0.022$ and BR$(\Xi^+_c\to p \bar{K}^*(892)^0)
/$BR$(\Xi^+_c\to p K^-\pi^+)
=0.54\pm0.09\pm0.05$.
\end{abstract}


\end{frontmatter}

Much less is known about the decays and branching fractions
of the charm baryons in comparison to the charm mesons;
even less is known about charm baryons containing strange
quarks. 
In this article, we confirm the existence of the Cabibbo-suppressed decay
of the $\Xi_c^+$ baryon:
$\Xi_c^+\to p K^-\pi^+$ (conjugate states should be
implicitly assumed throughout this paper), which was observed for the first
time only recently~\cite{Selex}.
We also provide new information
on the two-body decay $\Xi^+_c\to p\bar{K}^{*}(892)^0$.

FOCUS is a high energy photo-production experiment at Fermilab designed to study
charmed
particle physics. Charmed hadrons were produced by the interaction of a photon
beam
on a beryllium oxide (BeO) target.
  Different target setups were used
during the data taking;  about 2/3 of the data were collected using a
segmented BeO target interleaved with high resolution silicon
detectors (TS). The decay products were reconstructed using a large acceptance
multi-particle spectrometer. A vertex detector (SSD) composed of 12 planes of
silicon was used to provide high resolution tracking in the region immediately 
downstream
of the target, thus allowing the identification and separation of the charm
production and decay vertices.   Downstream of the SSD, tracking and momentum measurement were 
accomplished with five station of multi-wire proportional chambers and two large
aperture magnets operated with opposite polarity. Three multi-cell \v{C}erenkov
counters operating in threshold mode allowed identification of charged
electrons, pions, kaons, and protons
over a large momentum range. The spectrometer also contained a hadron
calorimeter, two electromagnetic calorimeters, and two muon detectors. 

Events were selected which contained at least 
one identified proton track, one kaon track, and one pion
track. The proton and pion had the same charge
while the kaon had an opposite charge.

Due to {\em{a priori}} likelihoods, the tightest \v{C}erenkov cuts were placed on the proton, the loosest cuts on
the pion. 
The proton candidate was selected by 
requiring that the proton hypothesis was favored (by the \v{C}erenkov analysis 
algorithm) over both the kaon and pion hypotheses by 0.7 and 4.0 units
of likelihood, respectively.  
The kaon candidate was selected by requiring the kaon hypothesis to be favored
over the pion hypothesis by 2.5 units of likelihood and not disfavoured
by the proton  
hypothesis by more than 1 unit of likelihood.
For the pion candidate, we required that the pion hypothesis was within 8.0
units of likelihood of the most likely hypothesis.
A more detailed discussion of the \v{C}erenkov analysis algorithm may be found 
elsewhere \cite{Wiss}.
All the candidate particles were also required to 
be inconsistent with a muon hypothesis using 
the muon detectors and also to be inconsistent with an electron-positron pair.
Furthermore kaon candidates were required to have a momentum greater than 
8 GeV/$c$. 

Known meson reflections where another particle was misidentified as the proton
were  reduced by raising \v{C}eren\-kov requirements on the proton candidate for
combinations compatible with the decay modes $(D^{+*}_s),D^{+},D_s^{+} \to K^+ K^- \pi^+$,
 and
$D^+\to\! K^- \pi^+ \pi^+$. A significant background contribution from
$\phi(1020)\to K^+K^-$ was suppressed in the same manner.

The process $\Xi_c^+\to \Xi^-\pi^+\pi^+$ has been selected as the normalization 
channel. 
For this decay pions
have been subjected to similar \v{C}erenkov cuts to reduce systematics,
while $\Xi^-$ hyperons were fully reconstructed via the channel
$\Xi^-\to \Lambda^0 \pi^-$. 
The $\Lambda \pi$ invariant mass was required to fall within three standard 
deviation of the known $\Xi^-$ mass.

After the candidate tracks were selected, the event was
 reconstructed using a candidate driven vertex algorithm which
is described in detail elsewhere~\cite{e687nim}. The algorithm forms a secondary
vertex from the candidate tracks and uses the total momentum vector as a seed
to construct a primary vertex by intersecting the seed with other tracks in the
event. The procedure returns the confidence level of the fits for the secondary
(CLS)  and primary (CLP) vertices 
plus the detachment significance between the two,
 $\ell/\sigma_{\ell}$, where $\ell$ is the
distance between the vertices and $\sigma_{\ell}$ is its error.
In this analysis, the $\ell/\sigma_{\ell}$ variable is the most powerful tool in
extracting the charm signal from the background.
The same algorithm also returns an estimator of the relative isolation
of the obtained  vertices. This estimator (ISO1) is the confidence level
that tracks  forming the secondary vertex might instead come from the primary
vertex.
The described set of variables  provides a good description of the topological 
configuration of the event such that an appropriate set of cuts on them effectively rejects the background.

We require $\ell/\sigma_\ell > 8$ as a compromise between the requirements 
to have good statistics and a good signal to noise  ratio.
The same vertexing cuts, CLS$\ge 1\%$, CLP$\ge 1\%$,
and ISO1$<95\%$ were used for both the Cabibbo  suppressed
channel and the normalization channel. 
For the Cabibbo suppressed decay we further require the proper time
resolution of the candidate to be less than 75~fs. Also, the primary vertex 
is required to have at least two tracks in addition to the charm seed. 

Using these selection  criteria, we obtain the mass plots for the 
$p K^-\pi^+$ and $\Xi^-\pi^+\pi^+ $ combinations shown in 
Figure~1~(a,b) respectively. In Figure~1~(a) the signal is clearly visible
in the inset while in the full view the 
$\Lambda_c^+\to p K^-\pi^+$ mode dominates.
The fit to the invariant mass distribution is performed with a Gaussian
function
for the $\Xi^+_c\to p K^-\pi^+$ signal
and a straight line for the background.
In Figure~1~(b) the fit to the $\Xi^- \pi^+ \pi^+$ invariant mass distribution is also
performed using a Gaussian signal and a linear background.  
The fits yield  
$202\pm35$ events for the $\Xi^+_c\to p K^-\pi^+$ channel and 
$133\pm14$ events for the $\Xi^+_c\to \Xi^-\pi^+\pi^+$ channel.
A cleaner sample for the $\Xi^+_c\to p K^-\pi^+$ can be achieved,
as shown in Figure~2, by means of tighter \v{C}erenkov cuts and a tighter 
requirement on the 
detachment significance $\ell/\sigma_{\ell}$. 

The efficiencies for the two decay modes were determined using a Monte Carlo
simulation.  The efficiency ratio for the two modes was found to be
$\varepsilon_{\Xi^-\pi^+\pi^+}/\varepsilon_{p K^-\pi^+}= 0.1539\pm0.0033$.
Correcting by this factor, the branching ratio is calculated to be
$0.234\pm0.047$ (statistical error only).

Extensive systematic studies and consistency checks were performed
on the branching ratio measurement. 
The behaviour of the  $\Xi^+_c\to p K^-\pi^+$ and $\Xi^+_c\to \Xi^-\pi^+\pi^+$ 
signals with each of the cuts used in the analysis was investigated and 
compared to the Monte Carlo predictions.
Also, split samples studies were performed which included separation into two 
different run periods, into particle and anti-particle, and into 
high and low momentum for the heaviest daughter particle. We checked that the
results were consistent for different topologies for the reconstructed $\Xi^-$
hyperons and for different histogram binning and fit conditions.
The results were compatible
within errors with the quoted branching ratio.
From all these studies, summarized in Figure~3 and in Table~1, we estimate an 
upper limit 
to the systematic error to be 0.022.

We have further investigated the resonant components of the
$\Xi_c^+\to p K^-\pi^+$ decay. Figure~4~(a) shows the Dalitz plot obtained
for a value of $\ell/\sigma_\ell >$10.
In the plot it is possible to recognize the presence of the $\bar{K}^{*}(892)^0$
meson. 
Due to the relatively high background level the resonant components have been 
evaluated from the sideband subtracted
two-body invariant mass spectra.
No structures compatible with $\Delta$ resonances have been observed (Figure~4~(b)),
while we observe structures of excited hyperons in the regions around 1.6 GeV/$c^2$ and at the edge of phase space around 2.25 GeV/$c^2$ as shown in Figure~4~(c).
However, they are not stable enough against variations in the analysis cuts and changes in the selection of sideband regions  for us to quote  branching fractions for them.
Thus we limited our consideration to the 
resonant component of the $\Xi_c^+\to p K^-\pi^+$
where the $K^-$ and $\pi$ form a $\bar{K}^{*}(892)^0$ meson (Figure~4~(d)).
The signal has been normalized using the value from the fitted three body 
invariant mass plot. All the fits were performed using Breit-Wigner 
distributions for the signals and linear backgrounds. the signal widths have 
been constrained to those predicted by the Monte Carlo.
The Monte Carlo simulation indicated a small (6\%) difference in efficiency 
between 
the resonant and non-resonant components of the decay which has been taken into
account.  
We determine the branching ratio to be:
$\Gamma(\Xi^+_c\to p \bar{K}^{*}(892)^0)*\mathrm{BR}(\bar{K}^{*}(892)^0\to K^- \pi^+)
/\Gamma(\Xi^+_c\to p K^-\pi^+)=
0.36\pm0.06$ (stat). Assuming Isospin conservation in the process $\bar{K}^{*}(892)^0\to K \pi$ we can 
correct this value for unobserved $\bar{K}^0 \pi^0$ decays, thus we get:
$\Gamma(\Xi^+_c\to p \bar{K}^{*}(892)^0)/\Gamma(\Xi^+_c\to p K^-\pi^+)=0.54\pm0.09(\mathrm{stat})$.
The consistency of the result has been tested using different sets of tighter cuts
and different choices for the sidebands. Using the fit variant method 
on the obtained result we estimate an upper limit to the systematic error 
to be 0.03 when only charged decays are considered and 0.05 after the correction for the neutral process are taken into account.

In conclusion we determine the branching ratio of the Cabibbo suppressed 
channel $\Xi_c^+\to p K^-\pi^+$ with respect to the Cabibbo favoured mode
 $\Xi_c^+\to\Xi^-\pi^+\pi^+$ to be:
$$\mathrm{BR} \frac{(\Xi_c^+\to p
K^-\pi^+)}{(\Xi_c^+\to\Xi^-\pi^+\pi^+)}=0.234\pm0.047 \mathrm{(stat)}\pm0.022
\mathrm{(syst)}.$$ 
This result is in good agreement with SELEX measurement~\cite{Selex}, $\mathrm{BR}(\Xi_c^+\to p K^-\pi^+)/\mathrm{BR}(\Xi_c^+\to\Xi^-\pi^+\pi^+)=0.20\pm0.04\pm0.02$.
We measure the relative branching ratio BR$(\Xi^+_c\to p \bar{K}^{*}(892)^0)/$BR$(\Xi^+_c\to p K^-\pi^+)$:
$$\mathrm{BR} \frac{(\Xi_c^+\to p \bar{K}^{*}(892)^0)}{(\Xi^+_c\to p K^-\pi^+)}=0.54\pm0.09\mathrm{(stat)}\pm0.05\mathrm{(syst)}$$
and detect the presence of ($K p$) hyperon formations, but no evidence of 
$\Delta$ isobars.

We wish to acknowledge the assistance of the staffs of Fermi National
Accelerator Laboratory, the INFN of Italy and the physics department
of the collaborating institutions. This research was supported in part
by the U. S. National Science Foundation, the U. S. Department of Energy,
the Italian Istituto Nazionale di Fisica Nucleare and Ministero
dell'Universit\`a e della Ricerca Scientifica e Tecnologica, the Brazilian
Conselho Nacional de Desenvolvimento Cient\'{\i}fico e Tecnol\'ogico,
CONACyT-M\'exico, the Korean Ministry of Education, and the Korean
Science and Engineering Foundation.

%
\begin{figure}
\epsfig{file=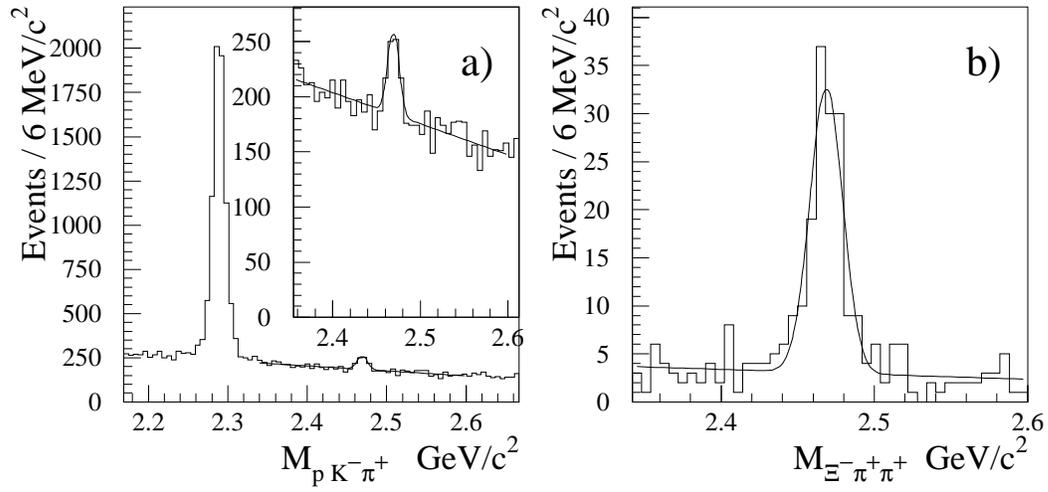,width=\columnwidth}
\caption{Invariant mass distributions: a)$p K^-\pi^+$; b)$\Xi^-\pi^+\pi^+$.} 
\end{figure}
\begin{figure}
\epsfig{file=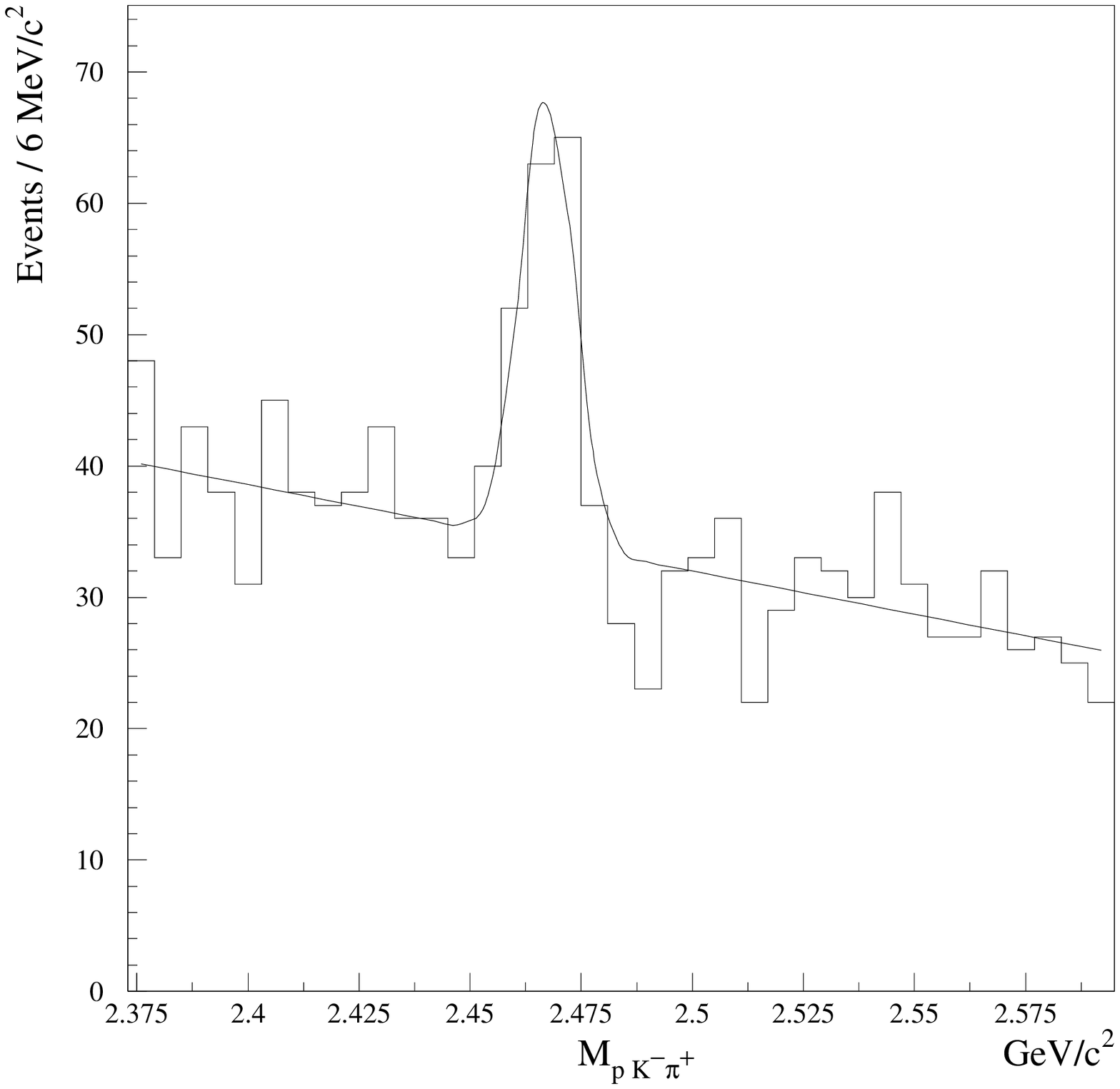,width=\columnwidth} 
\caption{Invariant mass distribution for the $\Xi_c^+\to p K^-\pi^+$ candidates}
\end{figure}
\begin{figure}  
\epsfig{file=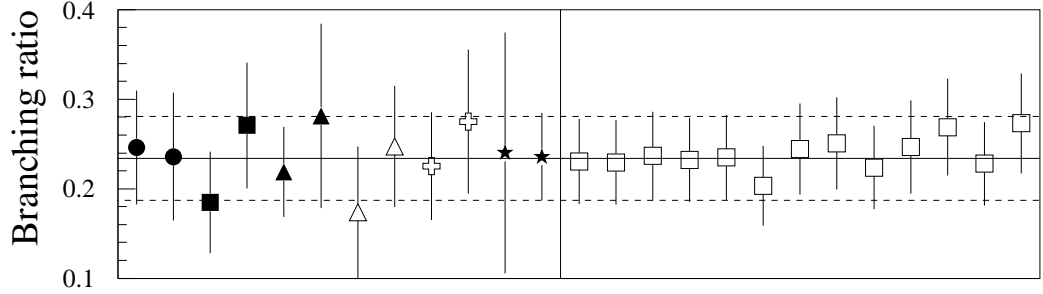,width=\columnwidth}
\caption{Systematic studies summary: split samples and fit variants.
First 12 points refer to the split sample test: high and low charm baryon momentum;
high and low heaviest daughter momentum; earlier and later runs; upstream and
downstream target region; particles and anti-particles; low and high value of
detachment significance $\ell/\sigma_\ell$. The other 13 points 
demonstrate the stability versus: different binning (points 13 through 16);
different fit intervals (points 17 through 25); and different background
functions (linear for points 17 through 19, second order polynomial for points
20 through 22, and third order for points 23 through 25.} 
\end{figure}
\begin{figure}
\epsfig{file=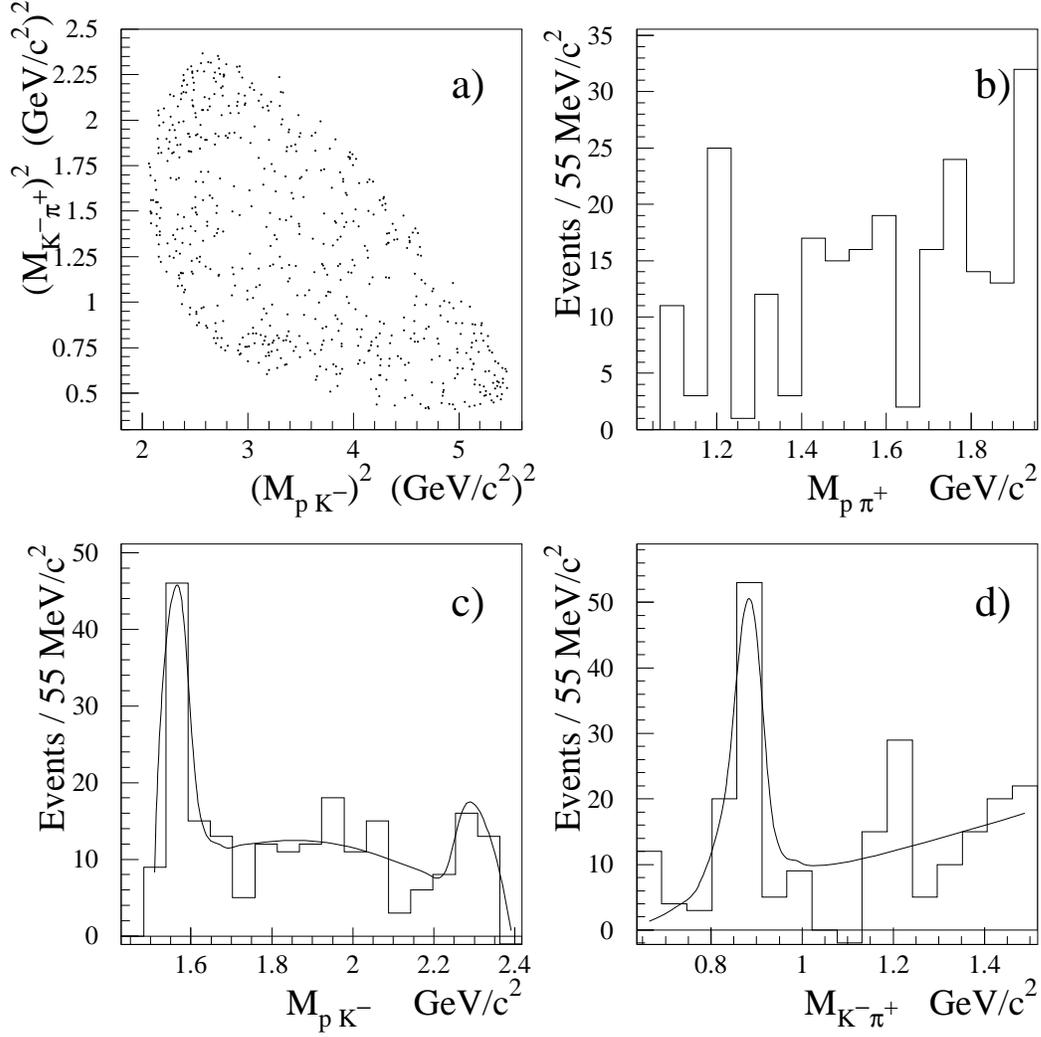, width=\columnwidth}
\caption{Analysis of resonant structure: a) Dalitz plot for the $\Xi^+_c\to pK^-\pi^+$ process; 
b) 2 body invariant mass distribution for the proton pion system;
c) 2 body invariant mass distribution for the proton kaon system;
d) 2 body invariant mass distribution for the kaon pion system.}
\end{figure}
\begin{table}
\begin{center}
\caption{Contributions to the systematic uncertainty for the measured branching ratio.}
\begin{tabular}{ll}
\hline
Split sample\rule{140pt}{0pt}& 0.0000 \\
$\Xi^-$ topology & 0.0018\\
Fit Variant& 0.0216\\
\hline
Total systematic error& 0.0223 \\
\hline
\end{tabular}
\end{center}
\end{table}

\end{document}